\documentclass[runningheads]{llncs}
\usepackage[colorlinks=true, linkcolor=blue, 
            citecolor=blue, urlcolor=blue,
            bookmarks=true,        
            bookmarksnumbered=true,
]{hyperref}
\usepackage{verbatim}
\usepackage[T1]{fontenc}
\usepackage{listings}
\usepackage{color}
\usepackage{graphicx}
\usepackage{tabularx}
\usepackage[colorinlistoftodos]{todonotes}
\usepackage[export]{adjustbox}
\usepackage[sort]{cite}
\usepackage{pbox}
\usepackage{siunitx}
\usepackage{xspace}
\usepackage{soul}
\usepackage{amsfonts}

\makeatletter
\renewcommand*{\thetable}{\arabic{table}}
\renewcommand*{\thefigure}{\arabic{figure}}
\let\c@table\c@figure
\makeatother

\begin{document}
\setcounter{secnumdepth}{3}
\newcommand\getcurrentref[1]{%
 \ifnumequal{\value{#1}}{0} {??} {\the\value{#1}}%
}    

\newcommand {\toolname} {{\sc ProPatrol}\xspace }
\newcommand {\hfocusobj} {{Active Execution Unit}\xspace}
\newcommand {\bfocusobj} {{Active execution unit}\xspace}
\newcommand {\mfocusobj} {{active execution unit}\xspace}
\newcommand {\hfocusobjs} {{Active Execution Units}\xspace}
\newcommand {\bfocusobjs} {{Active execution units}\xspace}
\newcommand {\mfocusobjs} {{active execution units}\xspace}

\newcommand {\focusobjs} {{Active-Execution-Units}\xspace }
\newcommand{\ctodo}[2]{\todo[inline]{\S\thesection:{\em #1}: {#2}}}
\newcommand{\vtodo}[1]{\ctodo{Venkat}{#1}}
\newcommand{\btodo}[1]{\ctodo{Birhanu}{#1}}
\newcommand{\rtodo}[1]{\ctodo{Rigel}{#1}}
\newcommand{\stodo}[1]{\ctodo{Sadegh}{#1}}
\newcounter{todoListItems}
\newcommand{\sstodo}[2][]
{\addtocounter{todoListItems}{1}
\todo[inline][caption={\hypertarget{todo\thetodoListItems}{}\thesection. #2}, #1]
{\hyperlink{todo\thetodoListItems}{#2} }}

\title{ProPatrol: Attack Investigation via Extracted High-Level Tasks}

\author{Sadegh M. Milajerdi\inst{1} \and
Birhanu Eshete\inst{2}$^,$\thanks{The second author performed this work as a postdoctoral associate at the University of Illinois at Chicago.} \and
Rigel Gjomemo\inst{1} \and \\
V.N. Venkatakrishnan\inst{1}}
\authorrunning{S. M. Milajerdi et al.}
\institute{University of Illinois at Chicago, Chicago IL 60607, USA 
\email{\{smomen2,rgjome1,venkat\}@uic.edu}
\and
University of Michigan-Dearborn, Dearborn MI 48128, USA \\
\email{birhanu@umich.edu}\\
}

\maketitle
\begin{abstract}
Kernel audit logs are an invaluable source of information in the forensic investigation of a cyber-attack. However, the coarse granularity of dependency information in audit logs leads to the construction of huge attack graphs which contain false or inaccurate dependencies.  To overcome this problem, we propose a system, called \toolname, 
which leverages the open compartmentalized design in families of enterprise applications used in security-sensitive contexts (e.g., browser, chat client, email client). 
 To achieve its goal, \toolname infers a model for an application's high-level tasks as input-processing compartments using purely the audit log events generated by that application. The main benefit of this approach is that it does not rely on source code or binary instrumentation, but only on a preliminary and general knowledge of an application's architecture to bootstrap the analysis.
Our experiments with enterprise-level attacks demonstrate that  \toolname significantly cuts down the forensic investigation effort and quickly pinpoints the root-cause of attacks. \toolname incurs less than 2\% runtime overhead on a commodity operating system.

\end{abstract}

\section{Introduction}\label{sec:intro}
Targeted and stealthy cyberattacks (referred to as Advanced Persistent Threats (APTs)) follow a multi-stage threat workflow ~\cite{mandiantapt1} to break into an enterprise network with the goal of harvesting invaluable information. APTs often utilize spear phishing and drive-by download to gain a foothold in an enterprise (initial compromise). After this step, APTs propagate to enterprise targets (e.g., Intranet servers) in pursuit of high-value assets such as confidential information.%

Once APTs are detected, it is crucial to track the causal linkage between events in a timely manner to find out the attack provenance. Consequently, attack provenance may be used to detect affected entities within a host or across multiple hosts. As soon as an attack provenance is uncovered, a system analyst can take immediate damage control measures, use it to make sense of past attacks or to prevent future attacks.

The state-of-the-art technique for provenance tracking is to use {\em kernel audit logs} to record information flow between system entities \cite{king2003backtracking,goel2005taser} and then correlate these entities for forensic analysis. In particular, after an attack is detected, system analysts use the detection point as a seed to initiate {\em backward tracking} strategies to determine the root-cause of that attack, and {\em forward tracking} methods to find out the impacts of the attack. 

Kernel auditing techniques interpose at the system call layer; therefore, they have acceptable runtime overheads but suffer from the \textit{dependency explosion} problem. In particular, due to coarse nature of dependencies that manifest in audit logs, an entity may {\em falsely} appear to be causally dependent on many other entities. For instance, consider a browser process that has multiple tabs open, each receiving data from different socket connections. If the browser process writes to a file, then during forensic analysis, that file will look causally dependent on all the socket connections the browser has accessed up to the write operation. In case of a drive-by-download attack that exploits that browser, it becomes challenging for system analysts to pinpoint the origin of the attack among all the accessed sockets.

To mitigate the dependency explosion problem, researchers have proposed compartmentalization techniques to partition the execution of a long-running process to smaller units \cite{lee2013high,ma2016protracer,ma2017mpi}. BEEP \cite{lee2013high} and ProTracer \cite{ma2016protracer} compartmentalize processes to low-level units based on iterations of event handling loops. MPI \cite{ma2017mpi} compartmentalizes processes to high-level tasks based on source code annotations manually performed by developers. Unfortunately, these techniques rely on source code or binary instrumentation.

{\bf Our Work.}
In this paper, we present an approach (called \toolname), aimed at high-level activity compartmentalization to address the dependence explosion problem and to provide {\em units of execution boundaries} to aid forensic attack investigation. One of the main benefits of our approach is that does not require application source/binary instrumentation.   The key insight in our approach is to leverage the  execution compartments that are inherent to the design of certain  Internet-facing applications (e.g., browsers, chat clients, email clients) in order to mitigate the dependence explosion problem during forensic analysis and are able  to pinpoint true dependencies. Through a {\em combination of execution compartments and provenance}, we demonstrate how a cyber-analyst can  perform precise forensic attack investigation.  Starting with the choice of compartmentalized applications, our approach also includes an inference mechanism to identify the execution compartments implemented in these applications directly from their audit log traces.

Our approach does require enterprise users to be restricted to the use of compartmentalized Internet-facing applications.  While this may seem stringent,  
recent trends \cite{win-locked-down} towards locked-down enterprise software (e.g., Windows 10 S) suggest that enterprises and software vendors desire this direction. In addition, modern applications are moving to a sandboxing-based architecture both for security and performance purposes. Google Chrome, for instance, is a relevant example of such an application, while Firefox is moving in the same direction~\cite{firefoxMulti}.

{\bf Results Overview.} Using APT scenarios as case studies, we evaluated the effectiveness and efficiency of \toolname on five attack scenarios in an enterprise setting. \toolname successfully constructed forensic graphs on five distinct lateral movement attempts that target high-value assets in Intranet servers. We note that these attacks cover a broad surface of the APT landscape. More precisely, our evaluation covers the major APT attack vectors such as spear phishing, drive-by downloads, and classic web-based attack vectors such as CSRF and DNS rebinding. 
In all the attacks, lateral movement is attempted by initiating a connection to an Intranet server. In addition to covering a wide space of APT vectors, our evaluation also spans web browsers, email clients, and instant messaging clients ---which are the common classes of applications targeted by APTs. Measured on the five attack scenarios for its runtime, on average, \toolname operates with an overhead of less than 2\%. Most importantly, \toolname is able to detect the execution compartments responsible for the attacks correctly in all the cases,  thus efficiently addressing the {\em dependency explosion} problem.

{\bf Outline.} The remainder of this paper is organized as follows. In Section \ref{sec:background} we motivate the problem by showing the importance of execution partitioning for better forensic analysis and describe details of Provenance Monitoring techniques that are required for log collection. Section \ref{sec:approach} discusses the details of our compartmentalization approach. In Section \ref{sec:impl}, we highlight implementation details. Evaluation of our approach appears in Section \ref{sec:eval}. Section \ref{sec:related} discusses related work. Finally, Section \ref{sec:concl} concludes the paper.

\section{Background}\label{sec:background}
\subsection{Motivating Example}

An enterprise network is typically composed of several employee machines and Intranet servers that host high confidentiality and high integrity assets. The network is often protected by a defensive perimeter consisting of firewalls and IDSs. In a typical setting, the employee machines may interact with external machines on the Internet, while the Intranet servers may receive connections only from inside the enterprise network. APTs typically exploit such connectivity of the employee machines to gain an initial foothold in the enterprise and subsequently perform lateral movement to reach high-value assets.

The most widely used APT attack vectors include sophisticated social engineering (e.g., spear phishing), browser compromises (e.g., via drive-by downloads), and web attacks (e.g., session riding)~\cite{johns2007protecting} that impersonate legitimate users of an enterprise host and connect to Intranet servers. Consider the following APT attack vector that highlights the need for precise provenance tracking.

Alice, an employee of an enterprise, has several tabs open on her browser. In one of the tabs, she is lured to a malicious website that contains a 0-day Java exploit that targets an unpatched Java plugin inside her browser. The exploit instantaneously drops an executable file, which is executed and spawns a shell where the attacker can remotely enter commands. Using this shell, the attacker reads Alice's recent activities from her command history and notices a series of \textit{git} commands to an internal GIT server. Next, the attacker executes a \textit{git pull} command to retrieve the most recent documents and proceeds to slowly exfiltrate them to a C\&C server that he controls. Alice is unaware of any of these actions.

This example showcases a drive-by-download APT attack vector\cite{attack-vectors}, a common method used to gain an initial foothold in an enterprise. The next step is typically gaining control of the compromised local machine followed by further connections to other internal machines. When a step of this attack is detected, it is crucial to causally link it with the events of the initial infection and ultimately with the provenance of the input that causes the initial infection. For doing so, we need to deal with several challenges pertinent to provenance tracking, dependence explosion, abstraction of input/output, dynamics of applications, and performance issues for timely analysis. 
 In particular, {\em dependency explosion} is one of the major hurdles to a fast and effective forensic investigation. This problem arises when a process receives several inputs 
from different sources within a short amount of time, while at the same time producing several outputs.
In this context, the primary challenge is to associate the provenance of each input to the correct outputs. 
For instance, the average number of records generated by the audit logs is typically between 5,000 - 500,000 records per minute, only a minuscule portion of which is related to the attack~\cite{hossain2017sleuth}.
\subsection{Provenance Monitoring}\label{sec:mon}

In this section, we describe details of a provenance monitoring system which produces logs required for building a dependency graph that is used for post-attack forensics analysis.
As the dependency graph is built based on information flow among system entities, we do not need to log all the system calls. Table \ref{syscall-types} shows a summary of the most important system calls that are required for information flow tracking and provenance identification. In the table, we show different categories of system calls according to their purpose. Some system calls are responsible for the actual information flow between objects. For instance, when a new process is created via a \texttt{clone} system call, it inherits the file descriptors of its parent. Therefore, there is information flow from the parent to the child process. 

A subset of the system calls (third row of Table \ref{syscall-types}) is responsible for initializing and setting up data structures rather than dealing with information flow directly. For example, the {\em socketpair} system call creates two sockets. \textit{Preparatory} system calls initialize data structures,  and in certain cases provide the provenance of the subsequent data. For example, by checking the {\em lseek} system call and considering file offsets, we only track specific offsets of a file to prevent unnecessary dependencies. \textit{Termination} system calls deal with the destruction of objects. 

\begin{table*}[!hbt]
\begin{tabularx}{\textwidth}{ | l | c | X |} 
  \hline      
  {\bf Purpose} & {\bf Relevant System Calls}\\
  \hline
  Information Flow  & {\em clone} (process), {\em fork} , {\em msgsnd}, {\em msgrcv}, {\em write}, {\em send}, {\em read}, {\em recv}, {\em exec}\\
  \hline
  Creation & {\em open}, {\em creat}, {\em dup}, {\em link}, {\em socket}, {\em socketpair} \\
  \hline
  Preparatory  & {\em lseek}, {\em connect}, {\em listen}, {\em accept}, {\em bind}, {\em clone} (thread), {\em link}, {\em sendto} \\
  \hline
  Termination  & {\em close}, {\em exit}, {\em exit\_group}, {\em unlink}, {\em kill} \\
  \hline  
\end{tabularx}
\vspace{1em}
\caption{System event types.}\label{syscall-types}
\end{table*}
\begin{table*}[!hbt]
\begin{tabularx}{\textwidth}{ | l | l | X | l | l |} 
  \hline      
 {\bf From} & {\bf To} & {\bf Relevant System Calls} & {\bf Source} & {\bf Destination}\\
  \hline
  Process & Process & {\em clone} (process), {\em fork}, {\em vfork}, {\em rfork}, {\em msgsnd} & event caller & arg(s) \\
  \hline
  Process & Process & {\em wait}, {\em msgrcv} & arg(s) & event caller \\
  \hline
  Process & File/Socket & {\em write}, {\em pwrite}, {\em writev}, {\em pwritev}, {\em send}, {\em sendto}, {\em sendmsg} & event caller & arg(s) \\
  \hline
  File/Socket & Process & {\em read}, {\em recv}, {\em recvfrom}, {\em recvmsg}, {\em execl}, {\em execv}, {\em execle}, {\em execve}, {\em execlp}, {\em execvp} & arg(s) & event caller \\
  \hline 
\end{tabularx}
\vspace{1em}
\caption{Information flow events.}\label{info-flow}
\end{table*}
{\bf Flow Types}. Table \ref{info-flow} shows the details of information flow sources, destinations, and events. We summarize these details in the table by using only three types of objects (File, socket, process).
As shown in the table, there are different kinds of information flow between system objects. These include: $(i)$ from a process to another process initiated by  events like {\em fork} and {\em clone},  $(ii)$ from a process to a file/socket initiated by events like {\em write} and {\em send}, and $(iii)$ from a file/socket to a process initiated by events such as {\em read}, and {\em receive}.
 In the last two columns of Table \ref{info-flow}, we use {\em arg(s)} to indicate the  argument(s) of system calls to refer to the object(s) that the {\em caller process} manipulates.
 In particular, depending on the system call, the argument type may be the {\em id} of a process, the {\em name} of a file, or a {\em descriptor} referring to a file/socket.

\section{Approach}\label{sec:approach}

\begin{figure}[!t]
  \centering
  \includegraphics[scale=.63]{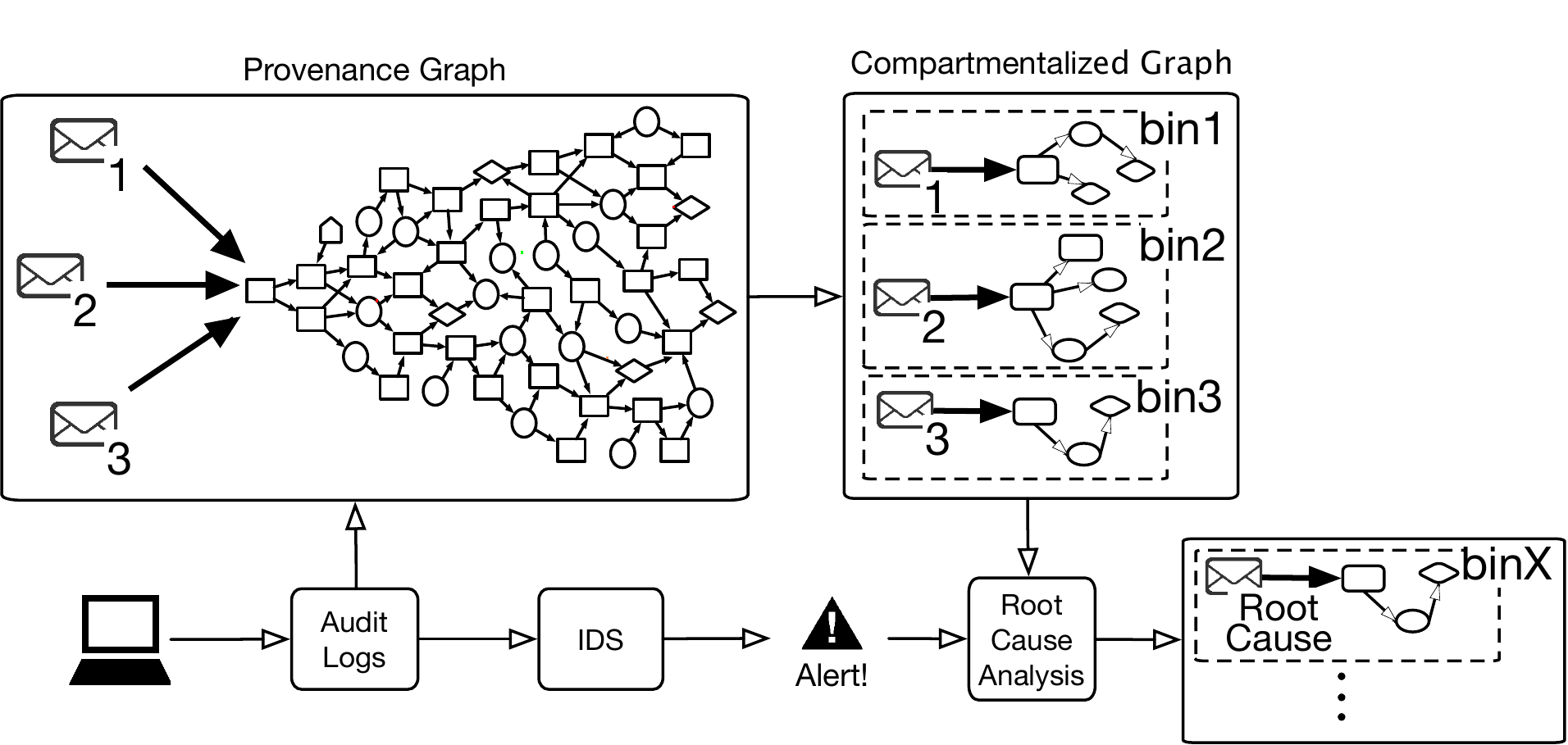}
   \caption{Approach Overview.}
     \label{fig:approach}
\end{figure}

{\bf Approach in a nutshell}. The goal of \toolname is to compartmentalize execution of long-running processes to smaller partitions by leveraging the high-level tasks extracted from audit logs. More precisely, after traces of an attack are detected, we want to perform forensic analysis and identify {\em who} initiated the connection (untrusted source versus legitimate local user), {\em how} it happened (history of the connection), and {\em what} system entities (processes, files, etc.) are affected. To answer these questions, we systematically follow the dependency between {\em system entities} (e.g., files, sockets, processes), which is constructed based on a system-wide provenance monitoring. This provenance monitoring is transparent to users, incurs negligible overhead, and does not require application instrumentation (details are discussed in \ref{sec:mon}). An overview of \toolname is shown in Figure \ref{fig:approach}. The provenance monitoring module constructs a dependency graph based on audit logs coming from enterprise hosts. Once an attack is detected, the compartmentalization module partitions long-running processes to smaller parts called \hfocusobjs, where each \hfocusobj relates inputs to outputs that are {\em truly dependent on those inputs}. For instance, in the case of a browser such as Google Chrome, each \hfocusobj represents a single user-supplied URL. Once the \hfocusobjs are determined, \toolname detects the root cause of the attack by performing a backward traversal from an attack point. In addition, it detects the affected objects by performing a forward traversal from the root cause. Thus, system analysts can quickly pinpoint the attack source and the affected system entities, which minimizes manual investigation efforts.

\subsection{Attack Investigation}

\paragraph{Solving the Dependency Explosion Challenge} In prior approaches dealing with \textit{dependency explosion}, a process is partitioned into smaller units of execution
based on performing heavy code instrumentation or assuming that source code is available and software developers annotate it~\cite{lee2013high, ma2016protracer,ma2017mpi}. They use each unit to next correlate the provenance of received inputs  to the produced outputs. On the contrary, we propose an approach that takes advantage of application compartments to learn a model through the analysis of the sequence of system calls it generates. Using this model, we define a partitioning scheme for applications, which assigns the provenance to the output objects of each \hfocusobj.

In particular, we define an {\em \mfocusobj} as the segment of an application that processes an input or a set of inputs as a result of user activity. Examples of such activities include reading a new email, browsing on a new website, and so on. We note that for the attacks that we deal with in this paper (e.g., drive-by download) we assume that such user activity is always present.

To use \hfocusobjs to assign provenance, we need to be able  to identify them from the system call traces when an \hfocusobj starts and when it ends. Besides, for every system call that interacts with an object (e.g., a file or a process) and that appears between that start and that end, we associate the provenance related to the \hfocusobj to that object. For instance, the \textit{\hfocusobj} of a browser such as Chrome is the website instance sending a request to Chrome's kernel process, while for an email client, the \hfocusobj is the email that the user is currently reading.

\subsection{\hfocusobj Identification}

Our methodology is based on an initial guided inference phase which exercises different applications with a variety of inputs. The inference is guided by an intuitive knowledge of what represents an \mfocusobj that might get compromised. For instance, for Google Chrome, \mfocusobjs are represented by visited websites, while for Thunderbird by the single emails. Such inference can be made with a high degree of certainty for several applications whose design and architecture are public knowledge, either because they are open source, or because of developer documentation.

When a new \hfocusobj starts, some system events are generated by the part of application that is responsible for handling a new \hfocusobj while others are commonly generated as a result of other application logic unrelated to this. Based on our observations, the latter represents a significant portion of the system calls and, during the inference phase, a source of `noise' for correctly deriving the boundaries between \focusobjs. Next, we propose a method to extract a sequence of system calls responsible for handling \hfocusobj.

Using our previous definition, we can partition an application into several \hfocusobjs, only one of which is active at any given time. To assign a system call to the correct \hfocusobj, we must, however, be able to identify which \hfocusobj is active when the system call is generated. 

In general, the problem can be defined as follows: Given a stream of system calls arriving one at a time, and given a set of bins, each representing an \hfocusobj, which is the active bin (that is the current active \hfocusobj), to which a system call belongs? 

We tackle this problem by creating an inference procedure, which exercises different activities to switch among \hfocusobjs and user actions that cause new input to be received. More formally, for each application, we want to derive the following rules during such inference phase:
\begin{itemize}
\item Rule 1: $ if(isObserved(S_k) \Rightarrow Bin_k=new Bin();)$
\item Rule 2: $ if(isObserved(S_i) \Rightarrow \{k=non\_active; i=active;\})$
\item Rule 3: $if(isActive(i))) \Rightarrow Bin_i = Bin_i \cup s_j$
\end{itemize}

Rule 1 deals with the creation of a new \hfocusobj (e.g., a new tab, or a new email which comes under the user's focus). In this rule, $S_k$ is a commonly observed sequence of system calls and their arguments when a new \hfocusobj is created, and $Bin_k$ represents a new empty bin. This sequence is typically manifested during the initialization of a new \hfocusobj. Rule 2 deals with the switching tasks among  different \hfocusobjs. In this rule, $S_i$ represents a commonly observed sequence of system calls when the user switches among \hfocusobj, $k$ and $i$ represent the previous \hfocusobj, which becomes inactive, and the newly activated \hfocusobj, respectively. Rule 3 deals with assigning the current system call $s_j$ to the currently active bin.

These rules are based on the key intuition that activities such as the creation of new \hfocusobj or switching among existing ones are executions of the same code in an application and they usually manifest in the same system call sequences.

To derive the sequences $S_k$ and $S_i$ for each application, we run that application under different scenarios (e.g., open a tab, click on a link in an existing tab or open the link in a new tab, or check an email or open an email in a new window, etc.), with different actions and user inputs. For each creation or switching, we record its start by introducing a special event (e.g., a mouse click) and collect the traces of system calls and their arguments, together with additional information such as PIDs and TIDs (thread ids). Next, we compare the sequences and extract the longest common subsequence among all the traces. 

More formally, given a set of system call traces, collected for the same type of activity repeated M times:
\begin{itemize}
\item $S_1= (s_{11}, s_{12}, s_{1N})$
\item $S_2= (s_{21}, s_{22}, s_{2N})$
\item ...
\item $S_M= (s_{M1}, s_{M2}, s_{MN})$
\end{itemize}
We find the longest subsequence $S_L = (s_{l1}, s_{l2}, s_{lK})$ where each $s_{li}$ is present in all the traces ($S_1, ..., S_M$), and where for any two consecutive $s_{li}$ and $s_{li+1}$ in $S_L$, $s_{li+1}$ follows $s_{li}$ in each of the traces ($S_1, ..., S_M$), possibly with other system calls between them. 
This subsequence represents a system call signature related to the specific activity, which is always present at the start of that activity. We use such subsequence as the `boundary' between the different \mfocusobjs.

After such subsequences are learned for a number of different activities under different inputs, we introduce them in the rules previously described. In particular, every time a new \hfocusobj's start is detected, we initiate a new bin. While this \hfocusobj is active, we assign the connections that are created to receive input to that bin. When a switch to an existing \hfocusobj is detected, we save the  state of the current \hfocusobj, in order to restore it once it becomes active again. If no subsequences can be identified, we conclude that the application is not suitable to be compartmentalized by our approach.

\section{Implementation}\label{sec:impl}
\graphicspath{ {pic/} }

\subsection{Provenance Monitor}
To trace the system calls, \toolname makes use of Systemtap~\cite{stap}, a very efficient Linux profiling tool designed to have near zero overhead. For each system call, we collect the timestamps, caller process id, group process id, system call name, and its arguments. We store these logs into a file to be analyzed further by the attack investigation module. Whenever a \texttt{fork} or \texttt{clone} appears in the logs, the new process or thread is monitored too.

\subsection{Compartmentalization}

\subsubsection{Google Chrome}~\\

To isolate websites from each other, Google Chrome consists of multiple renderer processes which communicate with Chrome's kernel process. Google Chrome supports different models of how to allocate websites to the renderer processes \cite{ChromeProcessModels}.  However, by default, it creates a separate renderer process for each web page instance which user visits. Each renderer process communicates the jobs to the kernel process and receives responses via the \texttt{recvmsg} system call.
Chrome's kernel process is responsible for networking and filesystem I/O tasks.  These jobs include DNS requests, content download, reading and writing to the file system and so on. 
Consequently, \toolname associates the provenance of an input with a renderer process which has sent a request to Chrome's kernel process.

\noindent
\textbf{\hfocusobj Selection.}
An \mfocusobj includes a renderer process and all the objects and processes initiated by the kernel in response to that renderer's messages. To be able to correctly assign and propagate the provenance, the attack investigation module must, therefore, associate each system call it receives with the correct \mfocusobj. We do this by taking advantage of the \texttt{recvmsg} system calls. In particular, when a \texttt{recvmsg} between the kernel and one of the renderer processes is found by the attack investigation module, we associate that system call, and all the subsequent system calls of the kernel and the renderer to the \mfocusobj related to that renderer. These system calls may include forking of new processes (e.g., plugins), writing to files, and so on. The new objects that are created or modified as a result of these system calls are associated with the provenance of the renderer. When a new \texttt{recvmsg} is `seen' by the attack investigation module from a different renderer process, we switch to the \mfocusobj corresponding to that renderer.

\subsubsection{Thunderbird}~\\

In the case of Thunderbird, each received email can be considered as a different sandbox associated with some provenance information related to the sender. Thunderbird stores all emails in a single file called INBOX and when a user opens a specific email, this file is accessed at an offset corresponding to that email using the \texttt{read} system call.

\noindent
\textbf{\hfocusobj Selection.}
An \mfocusobj in Thunderbird is defined as a set of objects to which information flows from Thunderbird as a result of reading an email. This set may include, files written by Thunderbird to the file system, browser processes forked by Thunderbird as a result of clicking on a link in an email and so on. In Thunderbird, each email is stored at a different offset in a single file, and Thunderbird uses this offset to access emails when prompted by the user. Therefore, when the attack investigation module finds a \texttt{read} system call into the INBOX file at a particular offset, it associates all the subsequent system calls with the \mfocusobj corresponding to the email at that offset.

\subsubsection{Pidgin Chat Application}~\\

Pidgin is a chat application. Each \mfocusobj in Pidgin corresponds to a chat window and the objects to which information flows from that window. Similar to Thunderbird, interaction with each chat window corresponds to access to a file. However, Pidgin keeps separate files for each chat window.

\noindent
\textbf{\hfocusobj Selection.}
\textit{Pidgin}'s screen is separated into different chat windows each of which corresponds to a different file on disk. Therefore, an \mfocusobj is switched by the attack investigation module when it finds a \texttt{read} system call to the file associated with a chat window.

\section{Evaluation}\label{sec:eval}
\graphicspath{ {pic/} }

\subsection{Enterprise Setup}

To evaluate the effectiveness of \toolname, we simulate a set of attacks on an enterprise testbed of user workstations and Intranet servers. In particular, the Intranet consists of three Ubuntu user workstations and  three Intranet servers. The Intranet servers include a GIT server used for collaborative coding within the enterprise, a web-based router, and a web server interfaced with a database that manages employees' personal information.%

\subsection{Graphs}
To facilitate forensic analysis, \toolname produces visual graph representations to be used by analysts.  In the Linux kernel, threads are implemented as processes that have the same process group. In the graph representation, we cluster the processes with the same process group (the process and its threads) together. The graphs depict processes and threads as ovals, sockets, and files, as well as information flow, labeled by numbers that show the sequence of events as they happened over time. Note that all the graphs we present in this section use these notations.

\subsection{Summary of Results}

Table \ref{tab:test-result}, summarizes \toolname's compartmentalization capability on highlighting five different classes of attacks that target common applications such as browsers and email clients. These attacks include Remote Administration Tools (RAT) installation via an attachment on a spear-phishing email, drive-by download that exploits a Java plugin vulnerability, social engineering via an IM client, CSRF, and DNS rebinding.  After the initial compromise in all these attacks, attackers pivot to one of the intranet machines that contain confidential information. At that point, an attack is detected, and system analysts use \toolname to find the root-cause of a connection to the corresponding sockets.

\begin{table}[!htb]
\centering
\small
\begin{tabular}{|l|l|c|}
\hline {\bf Application} & {\bf Attack} & {\bf Root-Cause Detection?} \\
\hline Email Client (Thunderbird) & RAT&\checkmark  \\
\hline Browser (Google-Chrome)& Drive-by download &\checkmark  \\
\hline IM Client (Pidgin) & Social engineering&\checkmark \\
\hline Browser (Google-Chrome) & CSRF&\checkmark   \\
\hline Browser (Google-Chrome) & DNS Rebinding&\checkmark  \\

\hline
\end{tabular}
\vspace{1em}
\caption{Overview of attack investigation results.}
\vspace{-4em}
\label{tab:test-result}
\end{table}
\subsection{Root-Cause Analysis}
Below we present details of the five scenarios on which we evaluated \toolname.

\subsubsection{Remote Access Trojan (RAT)} ~\\

{\bf Setup}. A RAT is a malicious binary that can execute several commands sent by the attacker. In this evaluation, we consider a spear-phishing email containing a RAT as an attachment. We assume that the user that receives this email is tricked into saving and executing the attachment. The attachment performs some malicious activity in the background without the user noticing it. In our evaluation, after it is downloaded to the user workstation, the RAT binary performs network scanning in the background. In this scenario, we used Nmap and a shell script that starts Nmap to mimic a RAT, which after being executed scans an internal IP address.

\begin{figure}
  \centering
  \includegraphics[width=.7\textwidth]{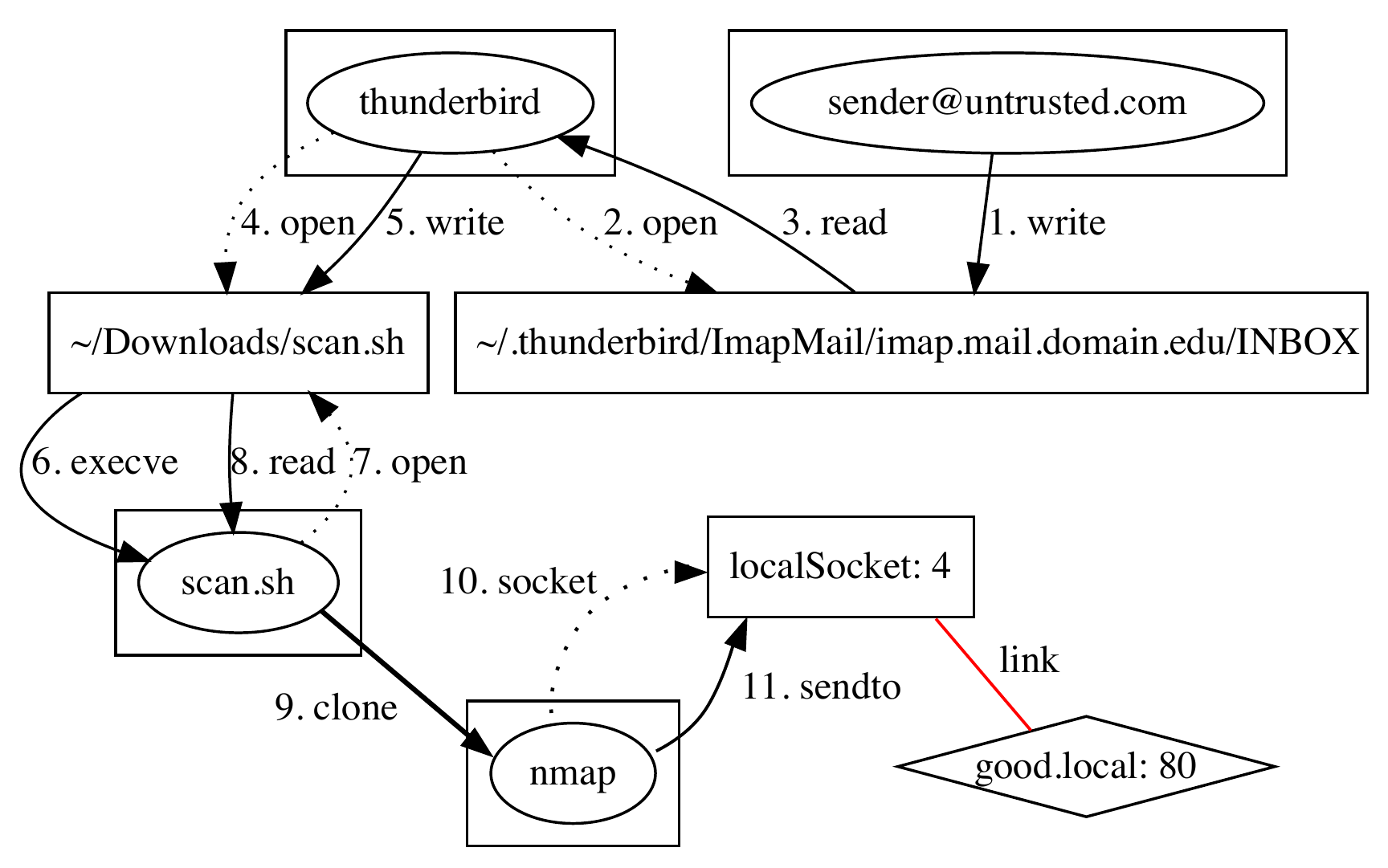}
    \caption{Provenance graph for RAT detection scenario.}
  \label{flow-RAT}

\end{figure}

 {\bf Attack Investigation}. Using \toolname, we were able to find the root-cause of this RAT starting from the connections sent to Intranet servers.
The sequence of system calls related to the  \mfocusobj is processed by \toolname to create a causal graph depicted in Figure \ref{flow-RAT}. Using this graph, a system administrator can easily infer
  the machine that has fallen victim to a malicious RAT that scans the network in the background.

\pagebreak
\subsubsection {Drive-by Download}~\\

{\bf Setup}. This attack exploits CVE-2012-4681, a vulnerability that allows a Java applet to bypass {\em SecurityManager} restrictions in Oracle Java Runtime Environment (JRE) version 7. We set up an external malicious web server that hosts a Java applet exploiting this vulnerability.  Whenever a victim browser with the Java Plugin connects to the malicious web server, the attacker can execute arbitrary code on the victim's machine. Specifically, we install JRE version 7 on the user workstations inside our network and set up the Java plugin for the Google-Chrome browser. Then we conduct the attack on one of the user's machines. The attack proceeds as follows. The user opens Google-Chrome, and among other benign activities, he opens a tab connecting to a malicious web server. When the user workstation connects to the malicious web server, the attacker notices this event. Then using Metasploit, the attacker opens a remote shell on the user workstation. 
Next, as a lateral movement for accessing the Intranet servers, the attacker tries to steal the enterprise project's data from the Git Server. Using the remote shell, the attacker performs \texttt{git pull } to pull the latest codebase of the project on the Git Server. Finally, the attacker sends the codebase to the attacker's server.

\begin{figure}
  \centering
  \includegraphics[width=1\textwidth]{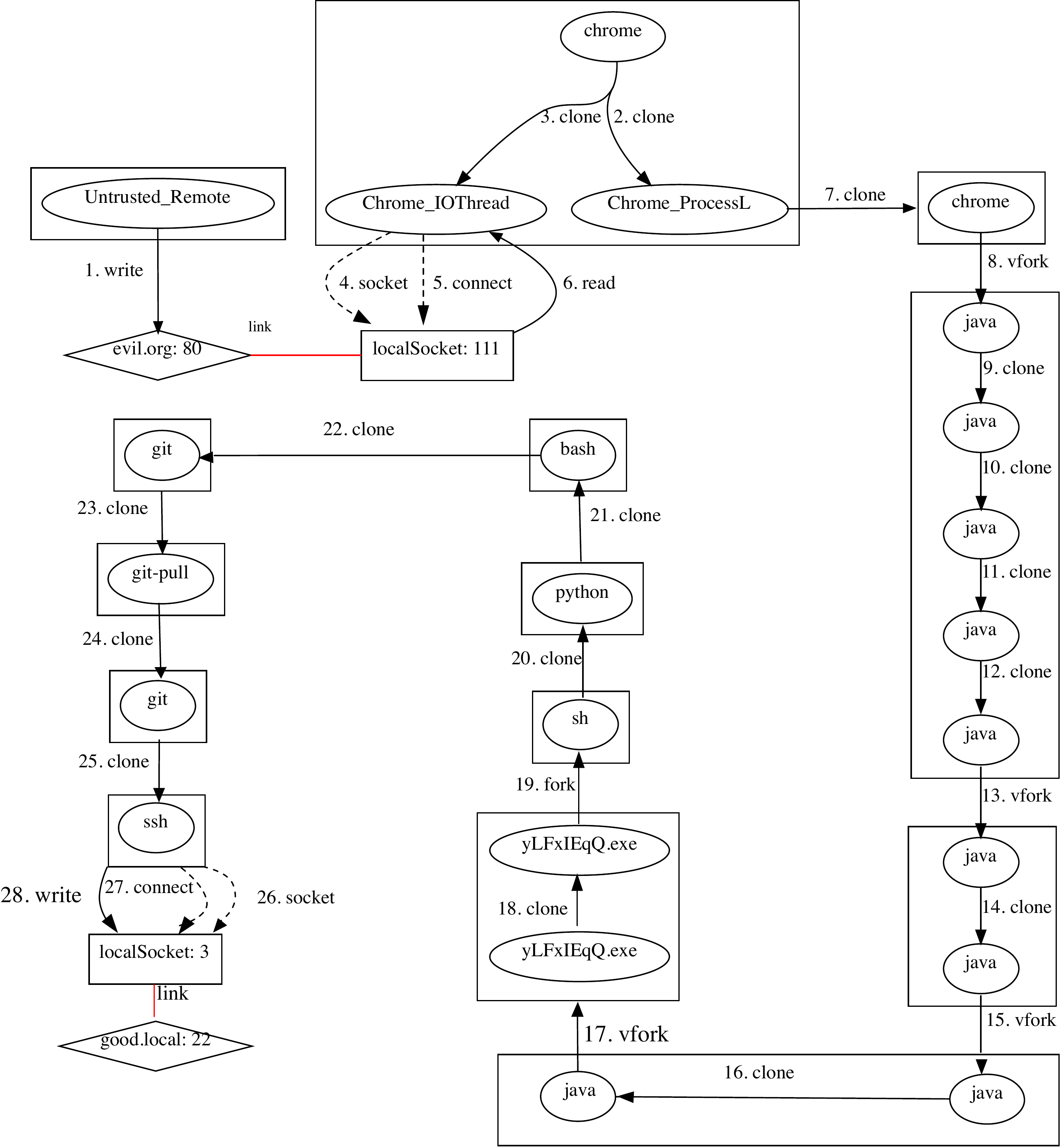}
   \caption{Provenance graph for drive-by-download attack detection.}
  \label{flow2}

\end{figure}

{\bf Attack Investigation}.  A provenance graph generated by \toolname, starting from a backward traversal from the \texttt{git} connection to the internal git server, is visualized in Figure \ref{flow2}. The first edge is artificial, and we consider it to show that the socket connection on port 80 of \texttt{evil.org} is on a malicious website. \texttt{Chrome} is the initial process that is executed by the user opening Google-Chrome. 
Later, that process clones two threads (\texttt{Chrome\_IOThread} and \texttt{Chrome\_ProcesssL}) (edges labeled with 2 and 3). The \texttt{Chrome\_IOThread} connects to the attacker's site and retrieves some data.
The thread (\texttt{Chrome\_ProcessL}) clones a set of different processes, threads, and applications for getting access to the remote shell (see edges 7-19). These intermediate steps are considered as internal mechanisms of Metasploit and the Java exploit we are using. After accessing the remote shell at edge 21, the attacker enters a \texttt{git} command. As the Git server uses SSH protocol, an SSH process is cloned and connects to port 22 on the Intranet Git server \texttt{good.local} (edges 26-28).

For the attack that exfiltrates the code base of the Git server, using the \toolname, we were able to identify the root-cause starting from the point that the attacker performs lateral movements to internal servers. 

\begin{figure}
  \centering
    \includegraphics[width=.7\textwidth]{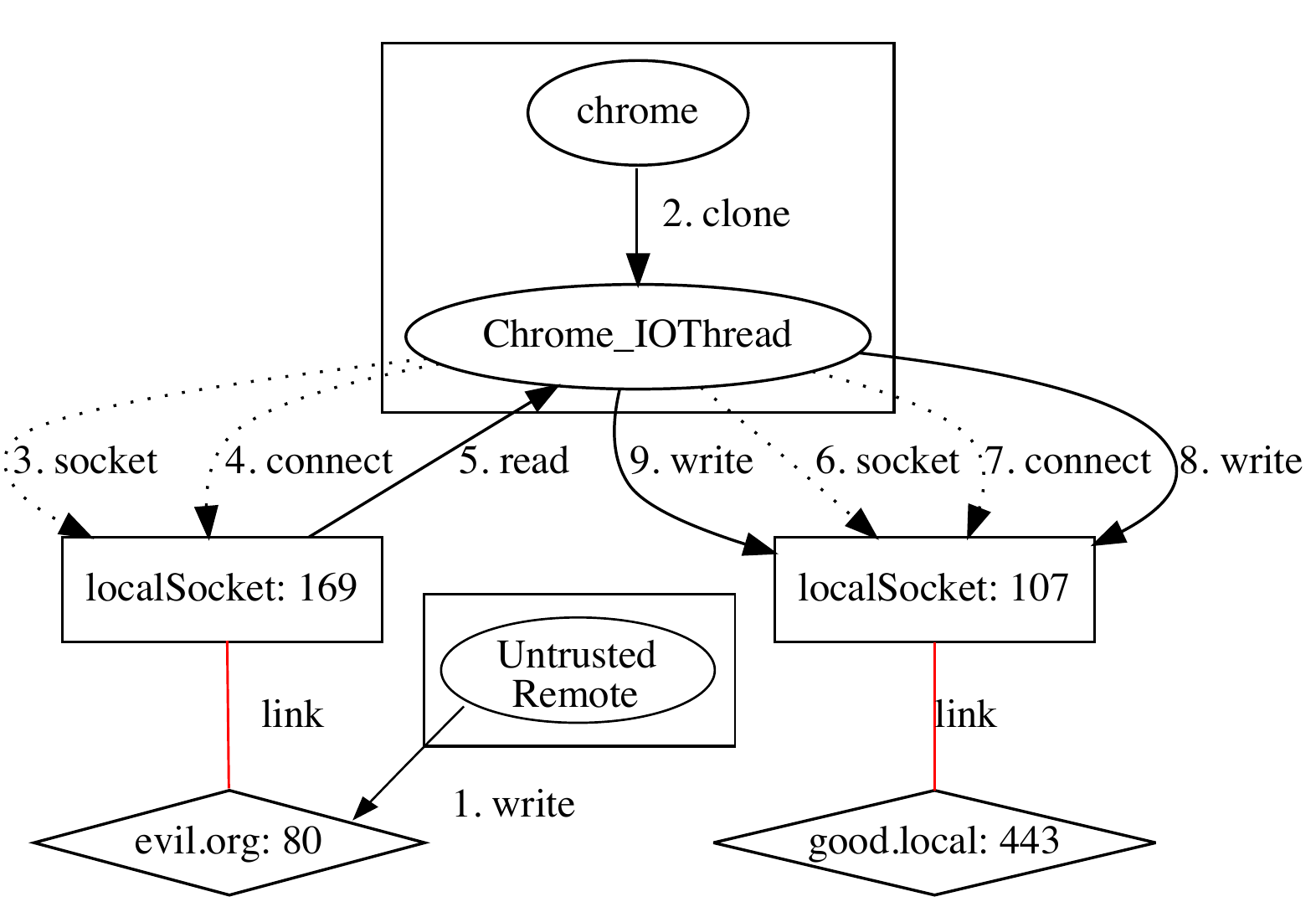}
      \caption{Provenance graph for web attack scenario involving CSRF and DNS rebinding.}
  \label{flow1}

\end{figure}
\subsubsection{CSRF and DNS Rebinding}~\\

In this class of attacks, we demonstrate how we investigate CSRF and DNS Rebinding using \toolname. We combined these two because of the similarity of the attack vectors.

{\bf Setup}. Our setup involves two malicious external web servers, one for CSRF and one for DNS Rebinding. The user workstation runs Google-Chrome with multiple open tabs. Some of the open tabs are connected to Intranet servers' sites. The user next uses one of the tabs to browse to one of the malicious websites, causing the browser to retrieve a page. Finally, the retrieved page sends a request to the Intranet server. 
For CSRF attack, we tested many different scenarios. These include: $(i)$ retrieving  a page that contains a hyperlink to an Intranet server and a user clicks on it to access the Intranet server, $(ii)$ retrieving a page that contains an element addressed by internal addresses and a JavaScript code snippet that checks the availability of those elements for port-scanning the enterprise network, and $(iii)$ a JavaScript code snippet sending a malicious GET/POST request to the webpage of the internal router having a CSRF vulnerability for changing the password. 

To evaluate \toolname against DNS Rebinding attacks, we set up a malicious external site containing a web server and a DNS server implemented with {\em Dnsmasq}. The DNS server has two IP addresses registered for the domain name of the web server, i.e., the IP address of the web server and the IP address of the Intranet server. When the user browser connects to this site and tries to resolve the domain name, the first IP it receives is the IP address of the web server. As a result, the browser connects to the web server and loads a webpage containing a JavaScript code that keeps requesting resources from the Intranet web server. These requests are blocked because Same-Origin-Policy prevents accessing the contents hosted on other origins. After a while, the user's browser connects to the DNS server one more time and tries to resolve the domain name again. This time, it is resolved to the Intranet web server's IP address, and the Same-Origin-Policy is circumvented ---enabling the attacker's script to read the response from Intranet server and send it to the attacker's machine.

{\bf Attack Investigation}. As shown in Figure \ref{flow1}, \toolname has detected the root-cause starting by a backward traversal from the attacker's attempt to send a malicious request to the Intranet server.
The first edge is artificial, and we consider it to show that the socket connection on port 80 of \texttt{evil.org} is on an untrusted site. \texttt{Chrome} process is the initial process that is executed by the user opening Google-Chrome. Later that process clones a thread named \texttt{Chrome\_IOThread}. This thread creates the local socket 169 and connects to the attacker's site. 
Then edge numbers 6 to 9 are events related to making a connection to the port 443 of the Intranet server \texttt{good.local}. Event numbers 8 and 9 transfer some untrusted information to the socket on the intranet server. %
Note that \toolname did not correlate the attack to the other valid requests going to Intranet servers in other applications or other tabs of the browser.

\subsubsection{Instant Messaging Client}~\\

{\bf Setup}. To demonstrate how \toolname forensically investigates attacks targeting Instant messaging clients, we considered the {\em Pidgin} IM client. Pidgin maintains individual conversation history in separate files for each contact of a user.
We add a google account in the Pidgin that contains a list of added buddies. Then we start chatting with some of them. For each buddy, there is a chat communication that is stored in a separate file from the other conversations.

\begin{figure}[!b]
  \centering
    \includegraphics[width=.7\textwidth]{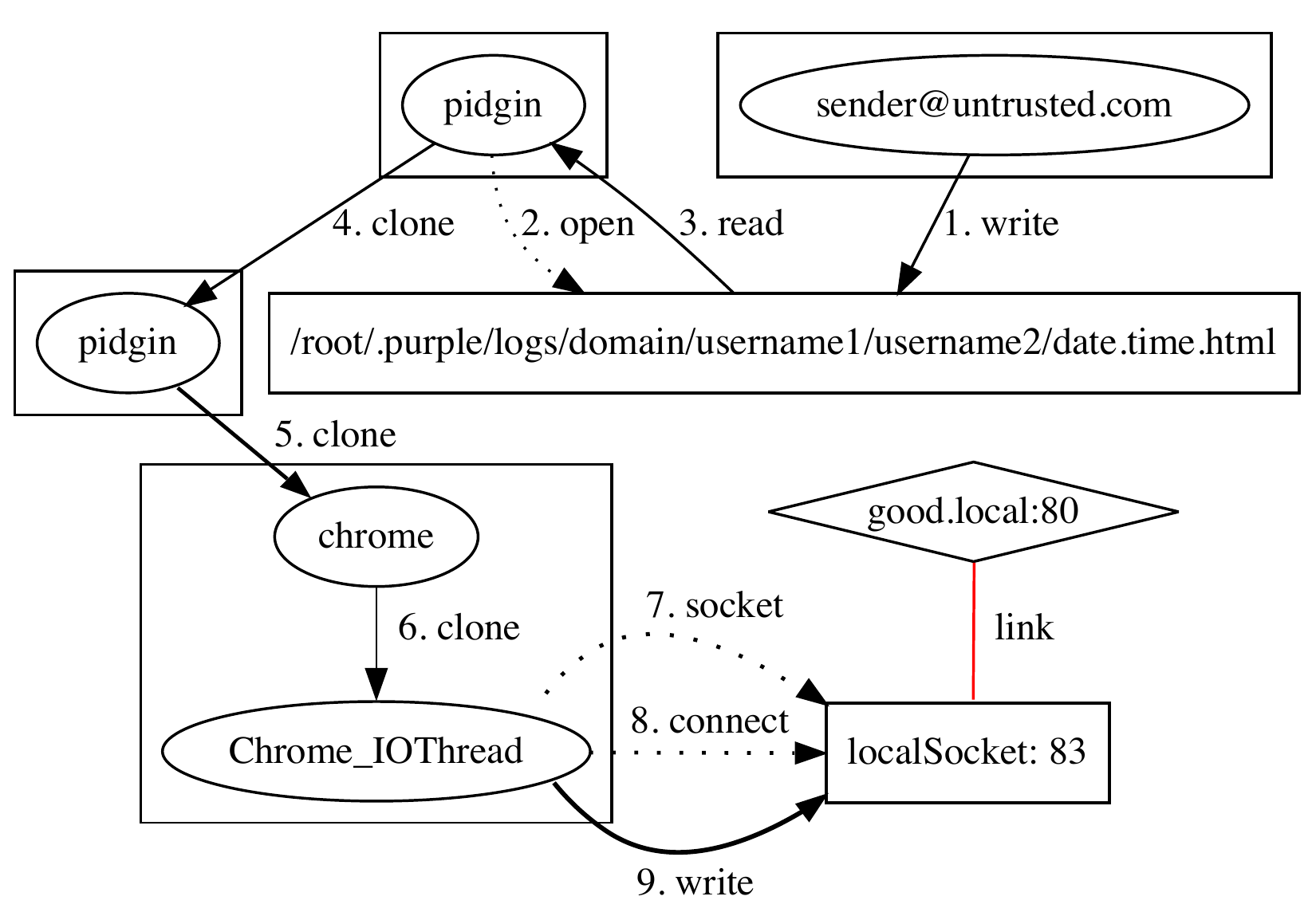}
      \caption{Provenance graph for an attack scenario that targets an IM client.}
  \label{flow4}

\end{figure}

{\bf Attack Investigation}. The provenance graph for detecting an attack that happens via an IM client is shown in Figure \ref{flow4}.  
In the chat communication with username2, username1 receives a chat messages with a link to a vulnerable Intranet server. 
When username1 clicks on the malicious link,  a Google-Chrome process is forked by pidgin, and a  connection to the Intranet server is initiated. At this moment, the \mfocusobj corresponds to the chat window with username2. Therefore, \toolname detects username2 as the root-cause of this attack.

\subsection{Effectiveness} 
Table \ref{tab:effect} shows the effectiveness of \toolname pertaining to event volume reduction. In this table, the second column shows the number of system calls generated by each application for its initialization which is the duration from the start of the application until it loads completely. Execution of each one of these applications could be compartmentalized to smaller bins depending on user activities. For instance, a new bin is created when a user opens a new tab in Chrome, or opens a new chat window in Pidgin, or reads a new email in Thunderbird. The third column shows the number of events assigned to each bin on average. For example, in the case of Google Chrome, if a user opens 10 tabs, the size of provenance graph would be about $200K+10\times14K$. In any attack, typically only one bin is responsible for the attack, and \toolname successfully identifies it. The fourth column shows the final number of events that \toolname shows to the system analyst after detecting the root-cause. As evidenced by this table, \toolname can achieve orders of magnitude reduction in event volume. 

\begin{table}[!t]
\centering
\small
\begin{tabular}{|m{2.4cm}|>{\centering\arraybackslash}m{2.5cm}|>{\centering\arraybackslash}m{2.5cm}|>{\centering\arraybackslash}m{4cm}|}
\hline {\bf Application} & \textbf{Initialization Syscalls} &\textbf{Average Bin Syscalls}  & \textbf{Average \mfocusobj Syscalls}\\
\hline Google-Chrome & 200K & 14K & < 50 \\
\hline Thunderbird & 91.5K & 8K & < 20 \\
\hline Pidgin & 20.5K & 1K  & < 15 \\
\hline
\end{tabular}
\vspace{1em}
\caption{Effectiveness of attack summaries.}
\vspace{-1em}
\label{tab:effect}
\end{table}

\subsection{Performance Overhead} 
Table \ref{tab:overhead} shows the performance overhead introduced by \toolname and the time required for generating the attack graphs. 
As shown in the second column of Table \ref{tab:overhead}, we calculate the average time for a single system call per scenario in microseconds. The third column shows the overhead (in percentage) by the monitoring infrastructure (which includes Systemtap and the provenance graph building module), which on average is 1.1.\%. We set up the graph generation module on a 32 bit Ubuntu OS, Quad-Core 2.4 GHz Intel Xeon Processor with 10 GB RAM. The time (in seconds) this module took for highlighting the root-cause of an attack is shown in the fourth column of Table \ref{tab:overhead} showing a very minimal overhead. Overall, both the graph generation module and Systemtap incur negligible overhead due to the coarse-grained provenance tracking underlying \toolname.

\begin{table}[!t]
\centering
\small
\begin{tabular}{|l|c|c|c|}
\hline {\bf Scenario} & \textbf{\pbox{2cm}{\bf Avg Event time (\SI{}{\micro\second})}} &\textbf{\pbox{3.5cm}{\bf Provenance Monitor Overhead (\%)}} & \textbf{\pbox{3.5cm}{\bf Graph Generation time (sec)}}  \\
\hline RAT& 8.87 & 0.72 & 0.12  \\
\hline Drive-by download& 16.37 & 1.57 & 0.001  \\
\hline Social engineering& 9.87 & 0.16 & 0.0004  \\
\hline CSRF& 8.10 & 1.14 & 0.03 \\
\hline DNS Rebinding& 15.71 & 1.88 & 0.08 \\
\hline
\end{tabular}
\vspace{1em}
\caption{Overhead for the Provenance Monitor and Graph Generation Time.}
\vspace{-1em}
\label{tab:overhead}
\end{table}

\section{Related Work}\label{sec:related}
\subsection{System-wide Provenance Collection}
SPADE \cite{gehani2012spade} and PASS\cite{muniswamy2006provenance} are operating system level provenance systems. SPADE hooks into the audit subsystem in the Linux kernel to observe program actions whereas PASS intercepts system calls made by a program.  Both of these systems observe application events such as process creation and input/output, which is then used to find out the relationship between data sets. LineageFS\cite{sar2005lineage} modifies the Linux kernel to log process creation and file-related system calls in {\em printk} buffer. A user-level process reads this buffer periodically to generate lineage records.  Similar approaches to collect provenance are Hi-Fi\cite{pohly2012hi} and LPM\cite{bates2015trustworthy} ---these are kernel level systems that track the provenance of system objects. While they provide a secure and application-transparent way of collecting provenance, they do need provenance awareness at the application level in order to counter the dependence explosion problem. Moreover, SLEUTH \cite{hossain2017sleuth} and HOLMES \cite{sadegh2019holmes} use kernel audit logs for real-time attack detection and forensics, which could benefit from the light-weight compartmentalization approaches such as \toolname to improve accuracy. %

\subsection{Information Flow Tracking}
Some past work (such as \cite{suh2004secure,yin2007panorama}) proposed information flow tracking at processor-level with manufacturer support. Some others (e.g., \cite{kemerlis2012libdft,newsome2005dynamic} perform binary rewriting at runtime to instrument machine code with additional instructions that update shadow memory. Xu et al. \cite{xu2006taint} employ source code transformation by instrumenting C code with additional code that can handle flow tracking.  Being fine-grained techniques, they offer good precision in tracking the source of enterprise activity. However, all these approaches impose a high overhead. For instance, ~\cite{kemerlis2012libdft} imposes a 3.65x slowdown factor. Another line of work also uses techniques to decouple taint tracking from program execution \cite{kwon18mci,ming2016straighttaint,chow2008decoupling,ji2017rain}.

In the coarse-grained tracking front, Backtracker by King et al. \cite{king2003backtracking} is one of the first works in this area that introduced the notion of dependency graphs. The same authors extended Backtracker in \cite{king2005enriching} with support for multi-host dependencies, forward tracking and correlating disconnected IDS alerts. To reduce the size of audit logs, different methods \cite{xu2016high,lee2013loggc,hassan2018towards,217579} are proposed leveraging graph abstraction, garbage collection, or compactness techniques.

\subsection{Execution Partitioning}
Execution partitioning techniques are proposed for dividing the execution of long-running programs into smaller units, resulting in a better forensic analysis. 
BEEP is a closely related approach to \toolname. BEEP  is based on the notion of independent units whereby a long-running program is partitioned into individual units by monitoring the execution of the program's event-handling loops, with each iteration corresponding to the processing of an independent input/request. An essentially backward forensic tracing system, BEEP, is suitable for programs that tend to have independent loop iterations.
Ma et al. \cite{ma2016protracer} introduced ProTracer, a lightweight provenance tracing system that only captures system calls related to taint propagation. ProTracer records the history of objects by logging important events.  It utilizes an instrumentation technique called BEEP \cite{lee2013high} for partitioning an execution into smaller units.  
BEEP \cite{lee2013high} and ProTracer \cite{ma2016protracer} use training and code instrumentation to divide execution to multiple iterations of the main loop in a program. Another related work, MPI \cite{ma2017mpi} relies on users to annotate the application's high-level task structures to enable semantic-aware execution partitioning.

\section{Conclusions}\label{sec:concl}
In this paper, we presented \toolname, as a compartmentalization approach for doing more accurate and timely root-cause analysis. \toolname uses a lightweight provenance monitoring system to effectively perform forward/backward tracking. Our evaluation shows that the tracking system operates with a very minimal overhead of less than 2\%. We demonstrated in an enterprise setting that \toolname is able to detect the root-cause of a broad class of APT vectors such as spear phishing, drive-by downloads, RATs, CSRF, and DNS Rebinding attacks. 
\section*{Acknowledgements}
This work was primarily supported by DARPA (under AFOSR contract FA8650-15-C-7561) and in part by SPAWAR (N6600118C4035), and NSF (CNS-1514472, and DGE-1069311). The views, opinions, and/or findings expressed are those of the authors and should not be interpreted as representing the official views or policies of the Department of Defense, National Science Foundation or the U.S. Government.

\bibliographystyle{splncs04}
\bibliography{main}
\end{document}